\documentclass[conference]{IEEEtran}

\usepackage{color}
\usepackage{url}
\usepackage{graphicx}
\usepackage{subfigure}
\usepackage{amsmath,amssymb,amsthm,latexsym}
\usepackage{hyperref}
\usepackage{flushend}
\usepackage{etoolbox}

\hypersetup{%
pdftitle={Cache policies for cloud-based systems: To keep or not to keep},
pdfauthor={Nicolas Le Scouarnec, Christoph Neumann and Gilles Straub},
pdfkeywords={cache policies, cloud},
pdfstartview={FitH},
urlcolor=darkblue,
pdfborder=0 0 0,
bookmarksnumbered
}%

\usepackage{ifthen}

\newboolean{showcomments}
\setboolean{showcomments}{true}    
\definecolor{gray}{gray}{0.8}
\newcommand{\comment}[2][]{%
  \ifthenelse%
    {\boolean{showcomments}}%
    {\colorbox{gray}{\parbox{0.95\linewidth}{\emph{\normalsize\sffamily{\bfseries\/Comment\ifthenelse{\equal{#1}{}}{:}{~#1:}}~#2}\/}}}%
    {}%
}

\def\dd{\,\mathrm{d}}
\def\expect#1{\mathbb{E}[{#1}]}

\newenvironment{newtext}{\color{blue}}{\color{black}}
\renewenvironment{newtext}{\color{black}}{\color{black}}

\begin{document}

\title{Cache policies for cloud-based systems:\\ To keep or not to keep}

\author{\IEEEauthorblockN{Nicolas Le Scouarnec, Christoph Neumann, Gilles Straub}
\IEEEauthorblockA{Technicolor\\
Email: \{nicolas.le-scouarnec, christoph.neumann, gilles.straub\}@technicolor.com}
}

\maketitle

\begin{abstract}
In this paper, we study cache policies for cloud-based caching. Cloud-based caching uses cloud storage services such as Amazon S3 as a cache for data items that would have been recomputed otherwise. Cloud-based caching departs from classical caching: cloud resources are potentially infinite and only paid when used, while classical caching relies on a fixed storage capacity and its main monetary cost comes from the initial investment.
To deal with this new context, we design and evaluate a new caching policy that minimizes the cost of a cloud-based system. The policy takes into account the frequency of consumption of an item and the cloud cost model. We show that this policy is easier to operate, that it scales with the demand and that it outperforms classical policies managing a fixed capacity.
\end{abstract}

\section{Introduction}

A variety of systems rely on caches: CPU caches speed-up access to memory pages; hard-disk caches speed-up access to files; network caches (e.g., proxies, CDNs) optimize network traffic, load, cost and speed-up data access. In this paper, we consider the specific case of caches in cloud-based systems. Cloud-based systems extensively rely on cloud storage (e.g., Amazon S3) and cloud compute (e.g., Amazon EC2), which changes the way to consider and design caches.

Cloud-based caching consists in storing frequently accessed files on cloud storage services (e.g., Amazon S3); it uses cloud storage as a data cache. Such a cache is beneficial for files that would have been recomputed using cloud compute or retrieved from a private data center at each access without a cache.

There are two main differences between cloud-based caching and classical network caching. First, cloud-based caching does not impose a limit on the cache capacity; the size of the cache is virtually infinite. Second, cloud-based caching adopts a pay-per-use cost model. Generally, the cost is proportional to the volume of cached data times the time data is cached. There is also a cost of not caching an item, e.g., the cost of (re)computing an item using cloud compute or the cost of fetching the item from a remote location.

The above differences impact the design of {\em cache policies} which determine how to efficiently use cache resources. Classical caches implement {\em cache replacement algorithms}, such as Least Recently Used (LRU) or Least Frequently Used (LFU), that decide which item to evict from the cache and to replace with a new item. These algorithms try to maximize the hit ratio of the cache.  With cloud-based caches it is not necessary to systematically {\em replace} an item, since the cache capacity is neither fixed nor bound. Further, the objective of the cache policy is not necessarily to maximize the hit ratio. 

In this paper, we design and evaluate cache policies with the objective of minimizing the cost of a cloud-based system. At each access of an item, the item is either recomputed at the cost of cloud compute or stored at the cost of cloud storage.  To keep or not to keep in cache, that is the question we need to answer once an item has been generated.

We focus on time-based cache policies. Time-based cache policies calculate how long an item should stay in the cache irrespective of a cache size limit. We show that these policies are easier to operate than classical size-based policies managing a fixed capacity. Time-based cache policies also scale with the demand and outperform size-based policies.

We instantiate and motivate the above caching problem with a cloud-based personalized video generation and delivery platform; in this use-case, video sequences contain pre-defined ad-placeholders, typically billboards that are included in the video content. Ads are dynamically chosen according to the end-user's profile and inlayed into the appropriate placeholders of the video using cloud compute. The video is cut into chunks such that the video processing only occurs on relevant portions of the video. Of course, several users (with overlapping profiles) may be targeted with the same ad for the same movie. Instead of recomputing these personalized chunks for each user (called items in the rest of the paper), the platform can cache them using cloud storage.

In summary, we make the following contributions:
\begin{itemize}
\item We provide analytical models to design and evaluate time-based caching policies that minimize the cost of cloud-based systems. The models allow computing the expected cost of the system. 
We derive three cache policies that minimize cost. The first requires a priori knowledge of the future as it computes the cost-optimal choice for taking the decision of caching; this policy provides a lower bound on the systems cost. The second requires a priori knowledge of the item popularity distribution and the global user request arrival rate; this global policy is simple to implement, yet fails at distinguishing items.
The third policy is applied individually for each item and requires a priori knowledge of the demand rates for each item. As we will show, its cost can be approached in praxis, even without a priori knowledge of item demand rates.
We validate the analytical models using simulation.
\item We propose a caching policy that relaxes the need for a priori knowledge, making it easy to deploy and operate in practice.  Still, the algorithm approaches the performance of the optimal cache policy with a priori knowledge on the item demand rates. The proposed caching policy applies individually on each item independently of other items. The sizes, request rates and costs of other items do not impact the caching decisions for one particular item. Further, the caching policy adapts over time to evolving request rates or prices.
\item We evaluate the proposed cache policies using both synthetic and trace-based simulations. We use traces from popular Video-on-Demand systems (Netflix, Youtube and Daum). Our evaluation shows that the proposed algorithm approaches the cost of optimal cache policies and scales with increasing demand. The evaluation also demonstrates that the classical LRU cache policy is not adapted to cloud-based caching.
\end{itemize}


\section{Motivation and Background}
\label{ref:motif}

In this Section, we present a motivating example of a cloud service that benefits from cloud-based caching. We further provide some background on cache policies.

\subsection{Cloud-assisted Targeted Ad-inlaying}
\label{sec:system}

We consider targeted ad-inlaying for a Video-On-Demand platform. Ad-inlaying consists in inserting ads as textures or objects {\em within} a video, in contrast to classical video advertising that interrupts the video to show ad-sequences. With ad-inlaying, ads cannot be skipped as being part of the content itself. In movie productions, the insertion of synthetic objects is done during video post-production (e.g., to add visual effects or synthetic commercial products). It is a costly process and requires the final rendering to be verified by humans (to ensure that the synthetic insertion will not be detected by users).

Classical ad-inlaying during post-production 
has limited personalization capabilities. In this work, we consider {\em targeted} ads, i.e., ads personalized towards individual users. We revisit the ad-inlaying pipeline to make it more scalable: any time consuming and manual tasks are done offline and only once for each movie. Automated tasks are done on-the-fly during content distribution using cloud infrastructure.

Manual operations of the ad-inlaying pipeline include the identification of ad-placeholders. We restrict the complexity of inserted objects to flat textures (typically rectangles), well suited for billboards. Humans identify the placeholders for the textures during post-production and  verify insertion and rendering using a single example texture. This process generates metadata describing the location and characteristics of the ad-placeholders. During offline processing, content is split in chunks of closed Group Of Pictures of equal size; only chunks that contain ad placeholders are recomputed and personalized during distribution. The latter chunks are encoded under a mezzanine format, being characterized by a higher video profile to enable further video editing \cite{mauthe2010mpeg}.

Video content is delivered to end users using HTTP adaptive streaming (e.g., HLS \cite{pantos2012http}, MPEG DASH \cite{stockhammer2011dynamic,sodagar2011mpeg}), which sends a playlist of URLs to the end user's video player. This enables content personalization: every user can receive a different playlist, containing some targeted chunks. 

During content distribution, we use cloud compute to generate (on-the-fly) targeted chunks, called hereafter items. Generating an item consists in (i) decoding the corresponding chunk under the mezzanine format, (ii) inserting the ad in every relevant frame of the video sequence using the corresponding metadata, (iii) compress the resulted video sequence under a distribution format compatible with the delivery network, and (iv) serve it to the user and store the item in a cache if needed. When a user asks for a content, the system decides which ads to insert into each placeholder of that content for that particular user.
The generation of the item using cloud compute can be economized if the item is present in the cache.

\begin{figure}
\begin{center}
\includegraphics[width=0.8\columnwidth]{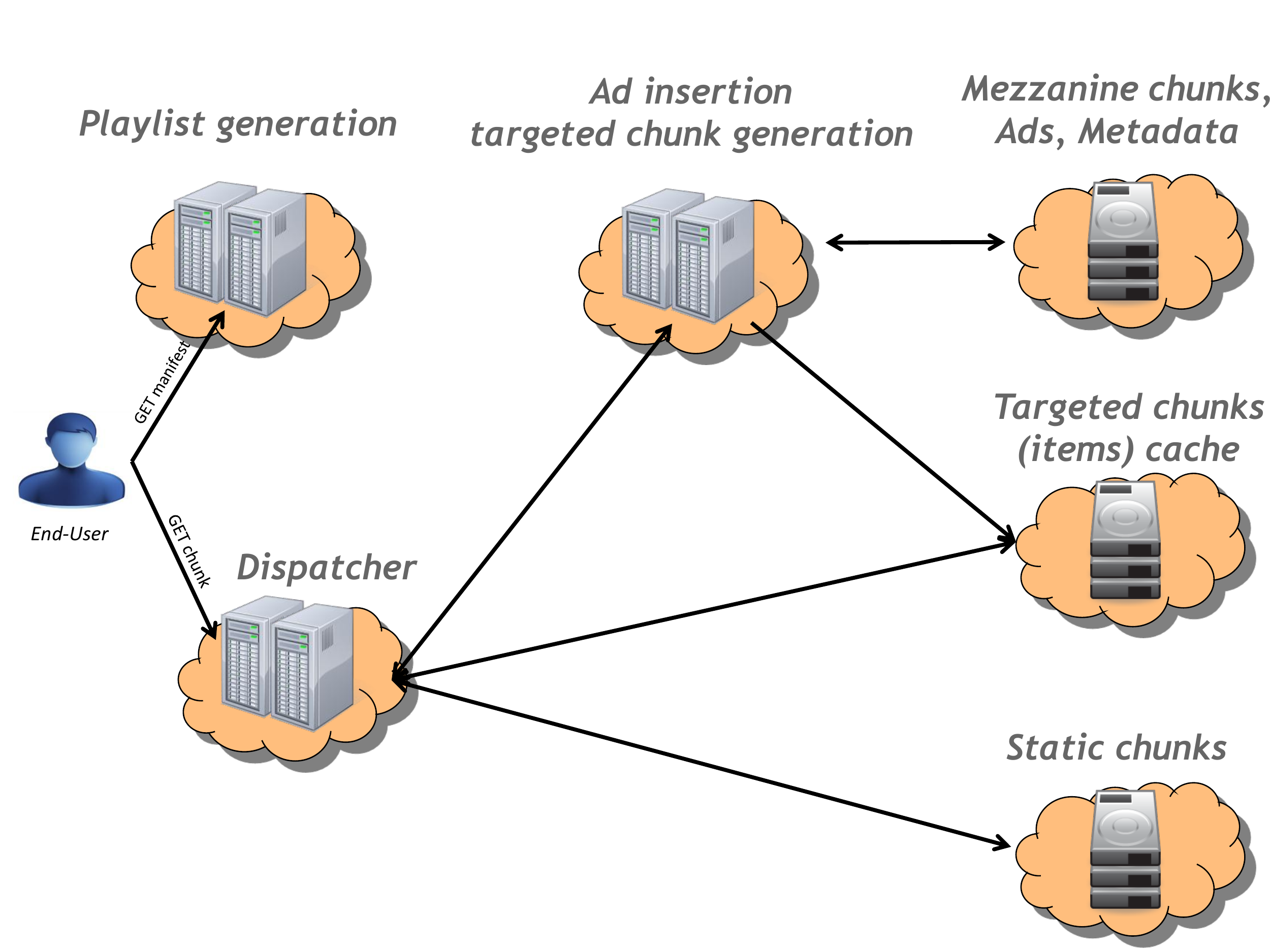}
\caption{Cloud-based targeted ad-inlaying architecture.}
\label{fig:Architecture}
\end{center}
\end{figure}

Figure~\ref{fig:Architecture} depicts a cloud-based targeted ad-inlaying architecture. 
Playlist generation, targeted chunk generation,  and the redirection of chunk requests are handled by cloud compute instances. Static chunks (non-targeted), mezzanine chunks and their metadata are stored on cloud storage. Targeted chunks (items) are cached on cloud storage using some caching policy.

A computing cost and a storage cost can be attached to any item while operating such a system on a cloud. In order to keep the cloud operating cost low, there is a need to define  the best strategy for either (re)computing or storing the items. This can be reformulated into a cache policy problem, that decides the items to keep or not to keep in the cache.
 
The same problem applies to other applications which require computation for media personalization. Nowadays, users get powerful TV sets with High Definition or Ultra high Definition and soon High Dynamic Range displays. Techniques such as upscaling or tone mapping are used to improve legacy content to get the best from these displays. However, their implementation residing in TV sets is often basic and could be either improved or personalized via the cloud computing power. The question ``to keep or not to keep'' remains valid and the strategies defined in this work apply.

\subsection{Background on Cache Policies}
\label{sec:background}
 
 
Cache policies have received a lot of research attention. Most caching policies assume fixed cache sizes and are flavors of LRU or LFU policies \cite{Megiddo2003,o1993lru,Cao1997,Rizzo2000}. An item is evicted from cache as a function of the item access frequency or the last time of access. These cache policies can be implemented in a cloud context by artificially limiting the cache size. However, as we will demonstrate these cache policies generally yield a higher cloud cost. In addition determining the cache size that would minimize the system cost is not practical, as it is sensitive to the system load (user arrival) and data popularity distribution. Analytical models for the LRU miss probability have been proposed \cite{Carofiglio2011,Jelenkovic2008}; these models do not consider the overall system's cost with a pay per-use cost model.

Time based cache policies apply a time-to-live (TTL) to every item stored in the cache \cite{Liu1997}.  The cache server decrements the TTL every second and evicts the item from the cache once the time-to-live reaches zero. Time based policies aim at maintaining data consistency. They do not impose a cache size limit by themselves but are often used in addition to some LRU or LFU policy that work with fixed cache sizes. Time based policies are used by DNS cache resolvers \cite{Sit2002} and web proxies \cite{Liu1997}. In general, the TTL value of an item is fixed by the origin server of the item (the authoritative DNS domain manager or the origin Web server). These policies do not try to optimize a cost with respect to a pay per use model.

The cloud infrastructure, by offering unlimited resources and a pay per use cost model, is an invitation to revisit cache policies. There is a need to define new cost effective policies that are not necessarily bounded by the cache size but rather consider the various cost factors.  Since time based cache policies by themselves do not impose a cache size limit, we focus on this type of policy in the remainder of this paper.

\section{Cache policies for cloud systems}
\label{sec:model}

In this Section, we suppose that requests for an individual item $k$ (corresponding to one chunk in a movie $i$ with an ad $j$) arrives according to an homogeneous Poisson process of intensity $\lambda_k$. We study the cost for serving a request when a given item is stored in a cache. The item is deleted from the cache if it is not accessed for more than $T_k$ seconds. If the item is not available from the cache it is computed. The storage cost is $S$ dollars per item per second, and the computation cost is $C$ dollars per item per computation. Note that, the cost of transmission and access counts can be included into costs $C$ and $S$. 

\def\dpzip#1#2{{{#1}^{s_#2}}{H_{N_#2,s_#2}}}
\def\pzip#1#2{\frac{1}{\dpzip{#1}{#2}}}

Let $t$ be a continuous variable describing the time since the last access to the item $k$. Since the request arrivals follow an homogeneous Poisson process, the probability that the next request for the item $k$ arrives at time $t$ is
\begin{align}
p(t) = \lambda_{k}\exp^{-\lambda_{k}t}
\end{align}

Let $X_k$ be a continuous random variable describing the cost for serving an item $k$. For a given request, if $t < T_k$, then the item is served from the cache and there are only storage costs $X_k=tS$. If $t > T_k$, then the item is stored for $T_k$ seconds, and re-computed when accessed. Hence, the expected cost for serving the item $k$ is
\begin{align}
\expect{X_k} = \int_{0}^{T_{k}}{p(t)tS\dd{}t}   + \int_{T_{k}}^{\infty}{p(t)(T_{k}S+C)\dd{}t}
\end{align}
which simplifies to
\begin{align}
\expect{X_k} = \frac{S}{\lambda_{k}} + \frac{(\lambda_{k}C-S)\exp^{-\lambda_k{}T_{k}}}{\lambda_k}
\label{eq:cost}
\end{align} 

The expected cost $\expect{X_k}$ has a minimum for $T_k=0$ if $\lambda_k <\frac{S}{C}$, or a minimum for $T_k=\infty$ if $\lambda_k > \frac{S}{C}$. 

Assuming that we perfectly know $\lambda_k$, \eqref{eq:cost} allows us to determine an ideal caching policy: \emph{(i)} never cache the item $k$ if its arrival rate $\lambda_k$ is smaller than the ratio cloud storage cost over cloud compute cost ($\lambda_k < \frac{S}{C}$); \emph{(ii)} indefinitely cache the item $k$ if its arrival rate is greater than the ratio cloud storage cost over cloud compute cost ($\lambda_k > \frac{S}{C}$).

In practice $\lambda_k$ is not known. In the next sub-sections we consider various policies derived from these observations.

\subsection{No knowledge of $\lambda_k$, Global TTL}
\label{subsec:global}
We first consider a setting in which we only know general laws governing the popularity distribution of movies. To this end, we assume that requests are distributed across movies according to a Zipf distribution of exponent $s_m$ on $N_m$ movies \cite{Fricker2012,Cha2007}. The probability that a request is for a movie $i$ is
\begin{align}
p(M=i)=\frac{1}{i^{s_m}}\frac{1}{H_{N_m,s_m}}
\end{align}
 where $H_{N,s} = \sum_{i=1}^{N}{\frac{1}{i^s}}$ is the generalized harmonic number.
 
We also assume that ads are distributed across movies according to a Zipf distribution of exponent $s_a$ on $N_a$ ads. 
$N_a$, $s_a$ and $N_m$, $s_m$ denote the Zipf distribution parameters for ads and movies respectively.
The probability that a request is for an ad $j$ is
\begin{align}
p(A=j)=\frac{1}{j^{s_a}}\frac{1}{H_{N_a,s_a}}
\end{align}

We note $p_{i,j}$ the probability that the $i$-th movie is served with the $j$-th ad (\emph{i.e.}, $k=(i,j)$). We consider that the choice of movies and ads are independent; we have
\begin{align}p_{i,j}=P(M=i)\cdot{}P(A=j)
\end{align}
 As a result, 
\begin{align}
p_{i,j}=\pzip{i}{m}\pzip{j}{a}
\end{align}
For the sake of clarity, we will note $p_{i,j}$ to designate this quantity in the rest of the paper.

We assume that the requests for each item also follow a homogeneous Poisson process, so that each item $i,j$ is served at a rate 
\begin{align}
\lambda_{i,j}=\lambda{}\cdot{}p_{i,j}
\end{align}
Hence, if the next request for some item arrives at time $t$ 
and the next request for each specific item $i,j$ arrives at time $t_{i,j}$%
, we have 
\begin{align}
P(t_{i,j}=t) = p_{i,j}
\end{align}
which means that movies and ads are distributed according to the Zipf-laws previously defined.

\def\summa{\sum_{i=1}^{N_m} \sum_{j=1}^{N_a}}

Let $X$ be the cost for serving any item. We do not distinguish items and set a global timeout $T$ that apply to all items (\emph{i.e.}, $T_{i,j}=T$). In this case, the expected cost $X$ for serving an item is
\begin{align}
\expect{X} = \summa {p_{i,j}\left(\frac{S}{\lambda_{i,j}} + \frac{(\lambda_{i,j}C-S)\exp^{-\lambda_{i,j}{}T}}{\lambda_{i,j}}\right)}
\end{align}

\begin{align}
\expect{X} = \summa {\frac{S}{\lambda} + \frac{\left(p_{i,j}\lambda{}C-S\right)\exp^{-p_{i,j}\lambda{}T}}{\lambda}}
\end{align}

\begin{align}
\expect{X} = \summa {\frac{S(1-\exp^{-p_{i,j}\lambda{}T})}{\lambda} + p_{i,j}\cdot{}C\cdot\exp^{-p_{i,j}\lambda{}T}}
\label{eq:Global TTL}
\end{align}

This policy is simple to implement yet fails at distinguishing items, and requires that the distribution of requests among movies matches well-known Zipf law. In the following, we design a policy that adapts the strategy to each item popularity rather than having a single strategy for all items.

\subsection{Estimation of $\lambda_{k}$, Individual TTL}
\label{subsec:indiv}
In this second policy, the arrival rate for each item $i$ is estimated by counting the number of requests occurring over a sliding temporal window. Following the previous observation that items for which $\lambda_{i,j} <\frac{S}{C}$ should not be stored at all, and that items for which  $\lambda_{i,j} > \frac{S}{C}$ should be stored indefinitely, this policy compares the estimate of $\lambda_{i,j}$ to $\frac{S}{C}$ to choose between storing or not. The decision of storing or not is continuously revisited each time the observed $\lambda_{i,j}$ changes.

In this case, the expected cost $X$ for serving an item is such that $T_{i,j}=0$ for all items such that $\lambda_{i,j} < \frac{S}{C}$ and $T_{i,j}=\infty$ for all items such that $\lambda_{i,j} > \frac{S}{C}$.  If we assume that $\lambda_{i,j}$ is perfectly estimated and that movies and ads are distributed according to
Zipf-laws as described in the previous section, then the expected cost $X$ for serving an item is
\begin{align}
\expect{X} = \summa {p_{i,j}\left(\frac{S}{\lambda_{i,j}} + \frac{(\lambda_{i,j}C-S)\exp^{-\lambda_{i,j}{}T_{i,j}}}{\lambda_{i,j}}\right)}
\label{eq:c1}
\end{align}
which can be rewritten as
\begin{align}
\expect{X} = \summa p_{i,j}{\left(\frac{S}{\lambda_{i,j}}(1-\exp^{-\lambda_{i,j}{}T_{i,j}}) + C\exp^{-\lambda_{i,j}{}T_{i,j}}\right)}
\end{align}
Leveraging the fact that $T_{i,j}=0$ for all items such that $\lambda_{i,j} < \frac{S}{C}$ and $T_{i,j}=\infty$ for all items such that $\lambda_{i,j} > \frac{S}{C}$, we obtain
\begin{align}
\expect{X} = \summa p_{i,j}{\left(\mathbf{1}_{\lambda_{i,j} \ge \frac{S}{C}}\cdot{}\frac{S}{\lambda_{i,j}} + \mathbf{1}_{\lambda_{i,j} < \frac{S}{C}}\cdot{}C\right)}
\label{eq:c2}
\end{align}

These formulas provide the cost of the best policy achievable if we have a perfect knowledge of $\lambda_{i,j}$ (and not an estimate). This allows comparing the performance of the scheme we implement, which estimates $\lambda_{i,j}$, against a policy with a perfect knowledge of $\lambda_{i,j}$.

\subsection{Perfect knowledge of event occurrences, Lower bound}
\label{subsec:lower}
Finally, this is an ideal policy which provides a lower bound that is not attainable: this policy assumes that when an event occurs at time $t$, we know at which time $t'>t$ the item will be requested next (\emph{i.e.}, we can predict the future).

We first determine the expected cost for serving a given item $k$.
\begin{align}
\expect{X_k} = \int_{0}^{C/S}{p(t)St\dd{}t}   + \int_{C/S}^{\infty}{p(t)C\dd{}t}
\end{align}
\begin{align}
\expect{X_k} = \frac{S}{\lambda}\left(1-\exp^{-\frac{C\lambda_k}{S}}\right)
\end{align}

If we assume that movies and ads are distributed according to a 
Zipf-laws as described previously, then the expected cost $X$ for serving an item $k=(i,j)$ is
\begin{align}
\expect{X} = \summa {p_{i,j}\frac{S}{\lambda}\left(1-\exp^{-\frac{C\lambda_{i,j}}{S}}\right)}
\label{eq:c3}
\end{align}

\subsection*{}
\begin{newtext}
An interesting feature of these three time-based caching policies is that their modeling and analysis is simpler than traditional capacity-based caching policies. Items do not interact with each other and in particular are not evicted from the cache so that another item can be stored. 
However, such time-based policies are adapted only to platforms that provide variable and unbounded amount of resources. Such platforms are now available from cloud provider that have a pay-per-use model with potentially infinite amount of resources.
\end{newtext}

\section{Evaluation}
\label{ref:eval}

\begin{newtext}
In this section, we validate the three analytical models previously defined using a corresponding cache policy implementation. 
We further evaluate the average cost of the different cache policies. For well-defined Poisson arrival of requests we use the analytical models to evaluate the average costs. For trace-based realistic arrival of requests, we have to use the corresponding implementation. 
For the purpose of comparison, we also include the LRU policy in our evaluation.
\end{newtext}

\subsection{Methodology}

\begin{figure*}
\begin{center}
\subfigure[Global TTL, no knowledge of $\lambda_k$]{
\includegraphics[width=0.365\linewidth]{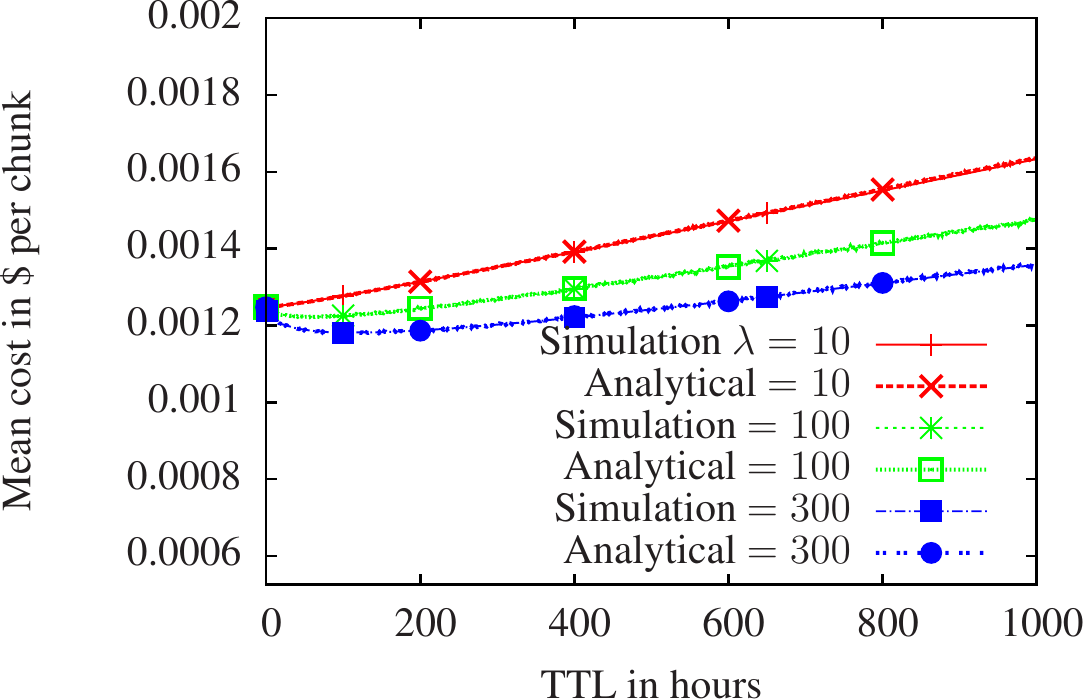}
\label{fig:ValidateGlobal TTL}
} %
\subfigure[Individual TTL, know./estim. of $\lambda_k$]{
\includegraphics[width=0.28\linewidth]{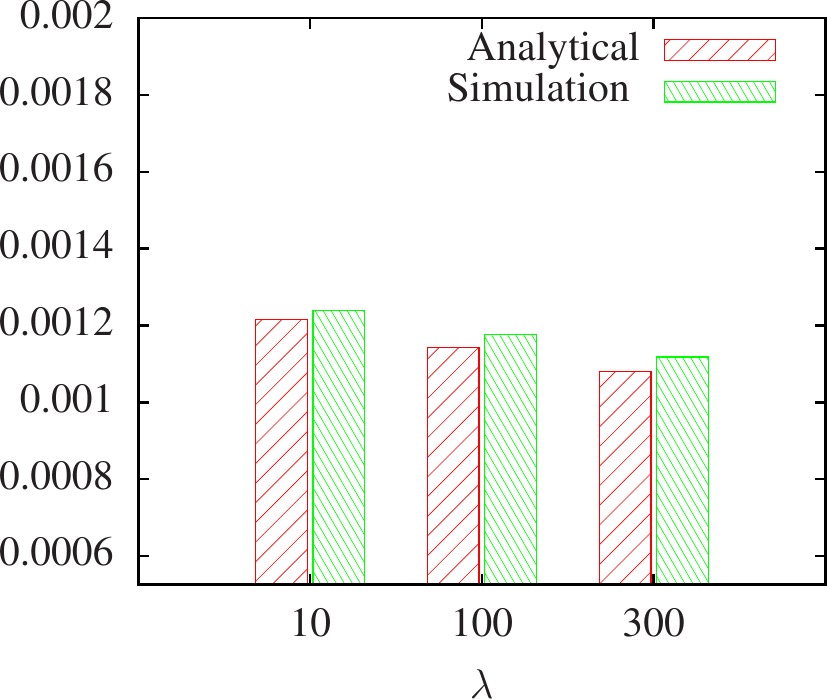}
\label{fig:ValidatePerItemOptimall}
} %
\subfigure[Lower Bound, knowledge of the future]{
\includegraphics[width=0.28\linewidth]{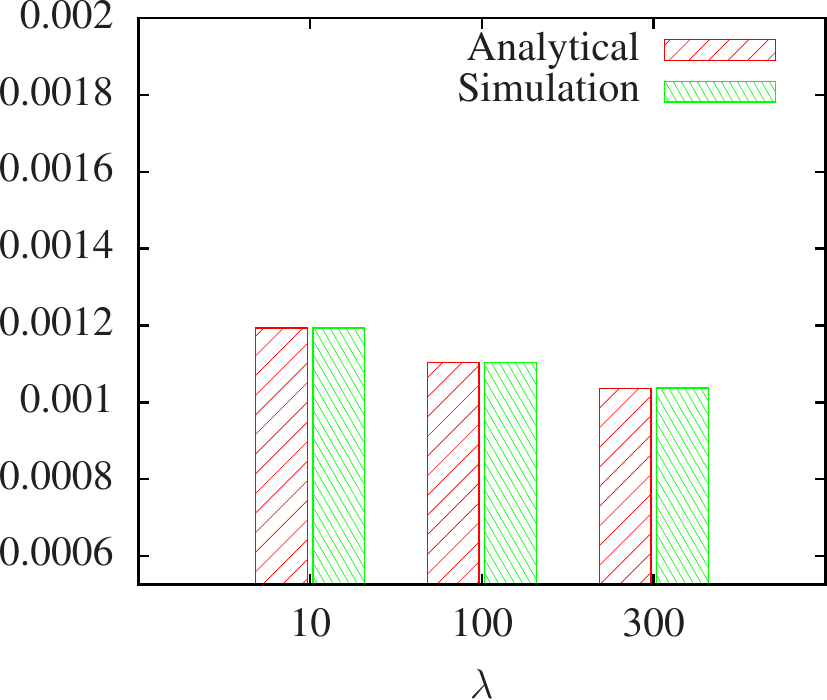}
\label{fig:ValidatePerfectKnowledgeOptimal}
} %
\caption{Comparison of the analytical model and the simulations. Mean cost per chunck for $\lambda = 10, 100, 300$, for $s_m=0.8$, $N_m=10000$, $s_a=0.91$, $N_a=500$ and for the various policies. The baseline on the y-axis (\emph{i.e.}, $0.000525$\$) corresponds to the fixed transmission cost.}
\label{fig:AnalyticalVsSimul}
\end{center}
\end{figure*}

We implemented a discrete-event simulator of a cloud-based caching system. The simulator models the arrivals of user requests, movie and ad popularities, and calculates the cloud system cost per item, given some cache policy. We use both synthetic and trace-based simulations.

Synthetic simulations use a Poisson arrival process to model request arrivals and two Zipf distributions to model movie and ad popularities. At each user request arrival, the simulator selects a movie $i$ according to a first Zipf distribution and an ad $j$ according to a second Zipf distribution. The tuple $(i,j)$ corresponds to an item as described in Section~\ref{sec:model}.
We use the values $s_m=0.8$ and $N_m=10,000$ \cite{Fricker2012,Cha2007} for the movie Zipf distribution, and  $s_a=0.94$ and $N_a=5,000$ for the ad Zipf distribution.\footnote{We extracted the latter parameter from an internal dataset of ad-popularity distribution of ads inserted on web pages.} We consider different global arrival rates $\lambda = 10$, $\lambda=100$, and $\lambda=300$ requests per hour.

Trace-based simulations rely on three different traces, representing three different types of Video-On-Demand systems: \emph{(i)} the Netflix prize dataset~\cite{Bennett2007}, representing a large premium content Video-On-Demand system, \emph{(ii)} Youtube traces~\cite{Cha2007}, representing a large user-generated content video system and \emph{(iii)} Daum traces~\cite{Cha2007}, representing a small user-generated content video system. 
The traces provide the simulator with user requests at various times: a user request corresponds to a movie view with Youtube and Daum or a movie rating with Netflix.
The Netflix dataset contains 17,770 movies and over 103 million movie ratings over a period of six years (2,237 days). The time of each rating is given in the trace (with granularity of a day). The Youtube trace is restricted to the Science and Technology category of Youtube, called Youtube Sci hereafter \cite{Cha2007}. The Daum trace is restricted to the travel category of Daum, called Daum Travel hereafter \cite{Cha2007}. The Youtube Sci and Daum Travel traces contain respectively, 252,000 and 9,295 movies, 539 million and 1 million movie views over a period of 570 and 160 days. The latter two traces provide the number of views at January 14th 2007 and the upload time of each video; we derive for each video the views per day, supposing a Poisson arrival process since the upload time of the video. To keep reasonable simulation times we reduced the Youtube Sci trace by randomly selecting 10\% of the items, obtaining a trace of 25,000 items, 53.9 million of views over a period of 570 days.
As with synthetic simulations, for each user request the simulator selects an ad according to a Zipf distribution with $s_a=0.94$ and $N_a=5,000$.

We derive our cost model from the cost model of Amazon AWS. 
We consider videos with an video output bitrate of $3.5$Mbit/s, which provides 720p HD quality. The video is split into chunks of length of $10$ seconds.
Our measurements indicate that we can compute an item in real-time using Amazon EC2 M1 Large instances \cite{bib:amazonec2instances}. At the time of writing these instances cost $0.26$\$ per hour for Europe \cite{bib:amazonec2pricing}. We can derive a cloud compute cost of $C = 7.2\!\times{}\!10^{-4}\,$\$ per item. Similarly, using the cloud storage costs of Amazon S3 \cite{bib:amazons3pricing} we can compute a storage cost of $S = 4.86\!\times{}\!10^{-7}\,$\$ per hour per item assuming a cost of $0.08$\$ per gigabyte per month.\footnote{Changing the quality of the video to 1080p HD or SD would not fundamentally change our results. Our measurements indicate that the ratio $S/C$ does not significantly change from one resolution to another.} 
\begin{newtext}
Finally, transmission costs are the same for cached and non-cached chunks, since Amazon only charges for traffic that leaves the Amazon cloud. In our model, the transmission cost accounts for $5.25 \!\times{}\!10^{-4}\,$\$ per item. In the rest of this section, all plots showing the average cost per item use $5.25 \!\times{}\!10^{-4}\,$\$ as their baseline for the y-axis.
\end{newtext}

The simulator implements the following cache policies:
\begin{itemize} 

\item {\em Global TTL:} simulates the policy described in Section~\ref{subsec:global}. It applies the same TTL to all items in cache. There is no cache size limit. The caching policy takes as parameter a TTL value.


\item {\em Individual TTL:} simulates the policy described in Section~\ref{subsec:indiv}.
It continuously adds or removes items to cache depending on an item arrival rate $\lambda_k$ observed during the sliding window on past item requests. With a $\lambda_k \leq \frac{S}{C}$ the simulator removes the item $k$ from the cache, and with a $\lambda_k \ge \frac{S}{C}$ the simulator adds the item $k$ to the cache.

We use a sliding window with a duration of 1,500 hours, which approximately corresponds to the value of $\frac{C}{S}$. We show in Section~\ref{sec:sliding} that a sliding window duration of $\frac{C}{S}$ provides best performance in most settings.

\item {\em Lower bound:} simulates the policy described in Section~\ref{subsec:lower}. It decides on each item request whether the item should have been kept in cache since the preceding request for the same item. The policy picks the cheapest of both choices. This computes a lower bound of the system costs for one particular simulator run. There is no cache size limit.

\item {\em LRU:} implements classical LRU policy with a fixed cache size. The caching policy takes as parameter a cache size.
\end{itemize}

\begin{newtext}
When evaluating analytically the cost of the policies for many values of $\lambda$ and many parameterization of the cache, it is not reasonable to completely evaluate the sum by iterating over all $i \in \{0\dots{}N_m\}$ with $N_m=10,000$ and all $j \in \{1\dots{}N_a\}$ with $N_a=5,000$). 
Given that all average analytical costs (see \eqref{eq:c1}, \eqref{eq:c2} and~\eqref{eq:c3}) are defined as:
\begin{align}
\expect{X}=\sum_{i,j} p_{i,j}\cdot{}\expect{X_{i,j}}
\end{align}
we approximate this computation using the Monte-Carlo method (25,000 random samples). The transmission cost per chunk is added to the cost obtained to plot curves.
\end{newtext}

\subsection{Model Validation}

We validate our analytical model by comparing the expected cost per item calculated by our analytical model with the cost per item provided by our simulator. Figure~\ref{fig:AnalyticalVsSimul} plots simulated and analytically calculated costs for {\em Global TTL}, {\em Individual TTL} and {\em Lower bound} policies. 

{\em Global TTL} (Figure~\ref{fig:ValidateGlobal TTL}) and {\em Lower bound} policies (Figure~\ref{fig:ValidatePerfectKnowledgeOptimal}) show a perfect fit between the simulator and the analytical model. There is a small difference for the {\em Individual TTL} policy (Figure~\ref{fig:ValidatePerItemOptimall}), due to the fact that in the simulator the arrival rate is measured over a sliding window, while it is exactly known in case of the analytical model. The simulator provides an estimation of the optimal cost by estimating the arrival rate, while the analytical model provides the bound.

Figure~\ref{fig:ValidateGlobal TTL} also shows that the TTL value minimizing the cost for the {\em Global TTL} policy depends on the global arrival rate $\lambda$. The optimal TTL is 0 (no storage at all) for $\lambda = 10$, and then increases with the value of $\lambda$.

\subsection{Comparing Size Based and Time Based Cache Policies} 

\begin{figure}
\begin{center}
\includegraphics[width=0.75\columnwidth]{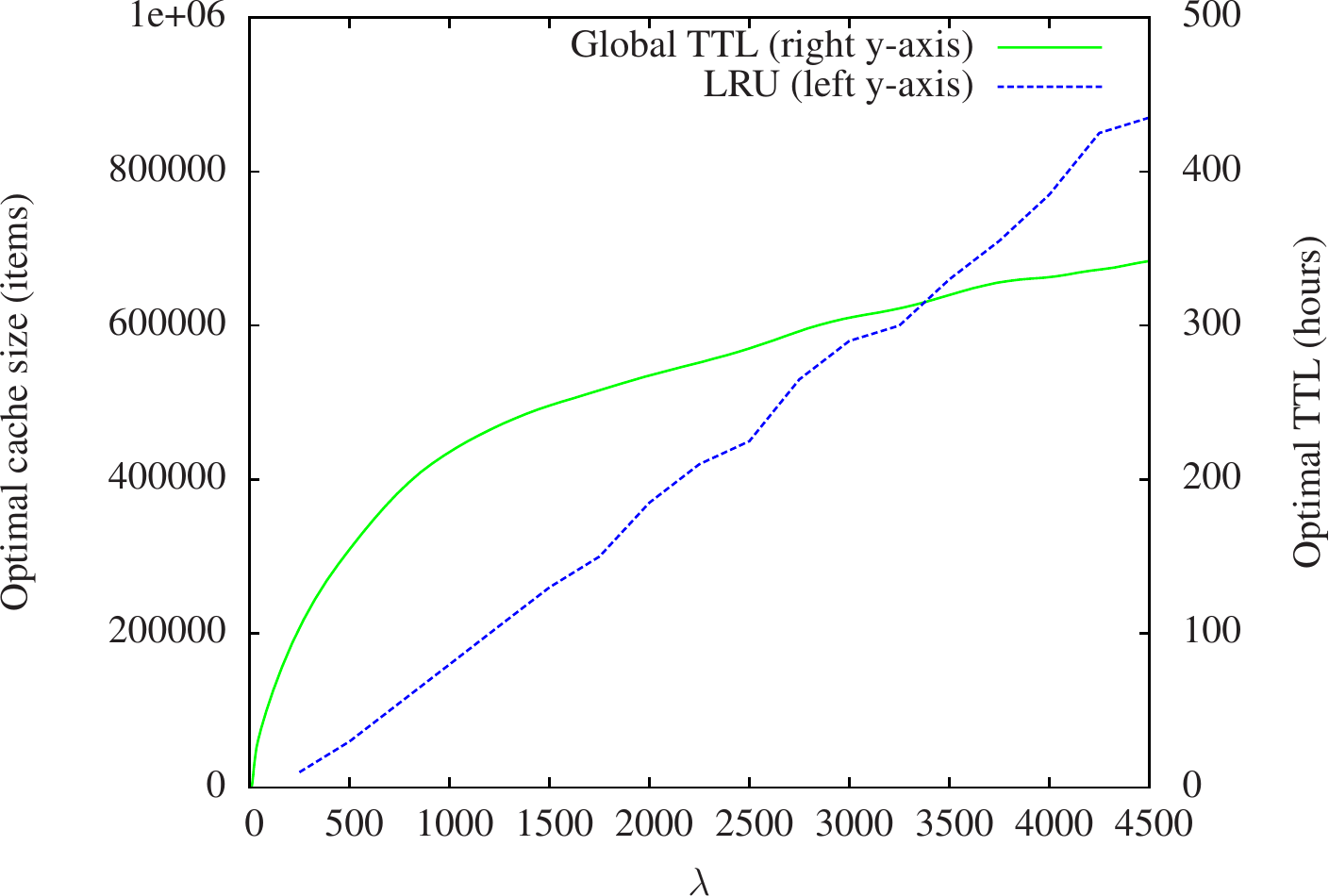}
\caption{LRU and Global TTL sensitivity to user request arrival rate.}
\label{fig:LRUTTLPerfComp}
\end{center}
\end{figure}

\begin{figure}
\begin{center}
\includegraphics[width=0.75\columnwidth]{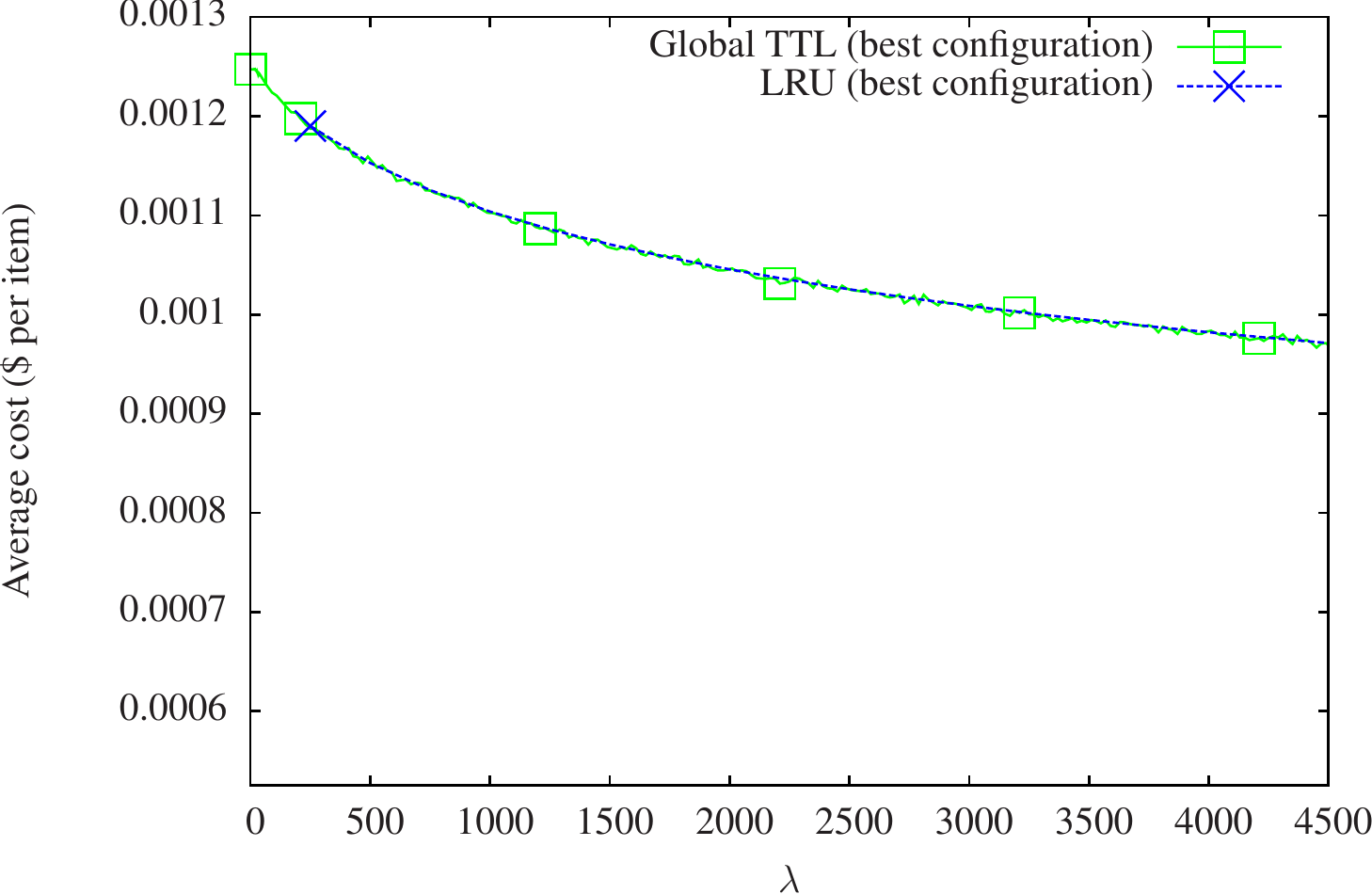}
\caption{Cost per item comparison of LRU and Global TTL, both parameterized at their best configuration (optimal cache size, and optimal TTL).}
\label{fig:LRUTTLCost}
\end{center}
\end{figure} 

We now compare a size based policy, namely {\em LRU}, and a time based policy, namely {\em Global TTL}.
We consider the cost per item for each policy and the sensitivity of the policy against increasing user request arrival rates.
We use the synthetic simulations for the LRU policy and the analytical model for the {\em Global TTL} policy. 
For both policies we determine the cache parameter (the size for {\em LRU}, the TTL for {\em Global TTL}) that minimizes the average cost per item for a given user arrival rate. It is not our objective to minimize the cache miss probability.
We use our simulator to compute the optimal {\em LRU} cache size in function of the user request arrival rate, and our analytical model to compute the optimal {\em Global TTL} in function of the user request arrival rate.

Figure~\ref{fig:LRUTTLPerfComp} plots the cache parameter that minimizes the cost as a function of the user request arrival rate for both policies. The {\em LRU} policy is more sensitive to the user request arrival rate than the {\em Global TTL} policy. The optimal cache size for the {\em LRU} policy increases as the user request arrival rate increases, while the optimal {\em Global TTL} seems to be bounded to a value less than 400 hours, as the user request arrival rate increases. 

Figure~\ref{fig:LRUTTLCost} shows the minimal cost of both {\em LRU} and {\em Global TTL} policies as a function of the user request arrival rate. The cost of both policies perfectly fit. The optimal {\em Global TTL} provides the same cost savings as the optimal {\em LRU} policy with less sensitivity to the user request arrival rate. Therefore, we do not consider the {\em LRU} policy in the remaining of our work and focus on time based policies ({\em Global TTL} and {\em Individual TTL}) and compare them against {\em Lower bound}.

\subsection{Time Based Cache Policies Evaluation}

\subsubsection{Using analytical models} 

We first compare the {\em Global TTL}, {\em Individual TTL} and {\em Lower Bound} policies using our analytical model. For the {\em Global TTL} policy, we determine the TTL value that minimizes the cost for this policy (the TTL minimizing {\em Global TTL} is 0, 60 and 120 hours for a $\lambda$ of 10, 100 and 300 respectively).

\begin{figure}
\begin{center}
\includegraphics[width=0.75\columnwidth]{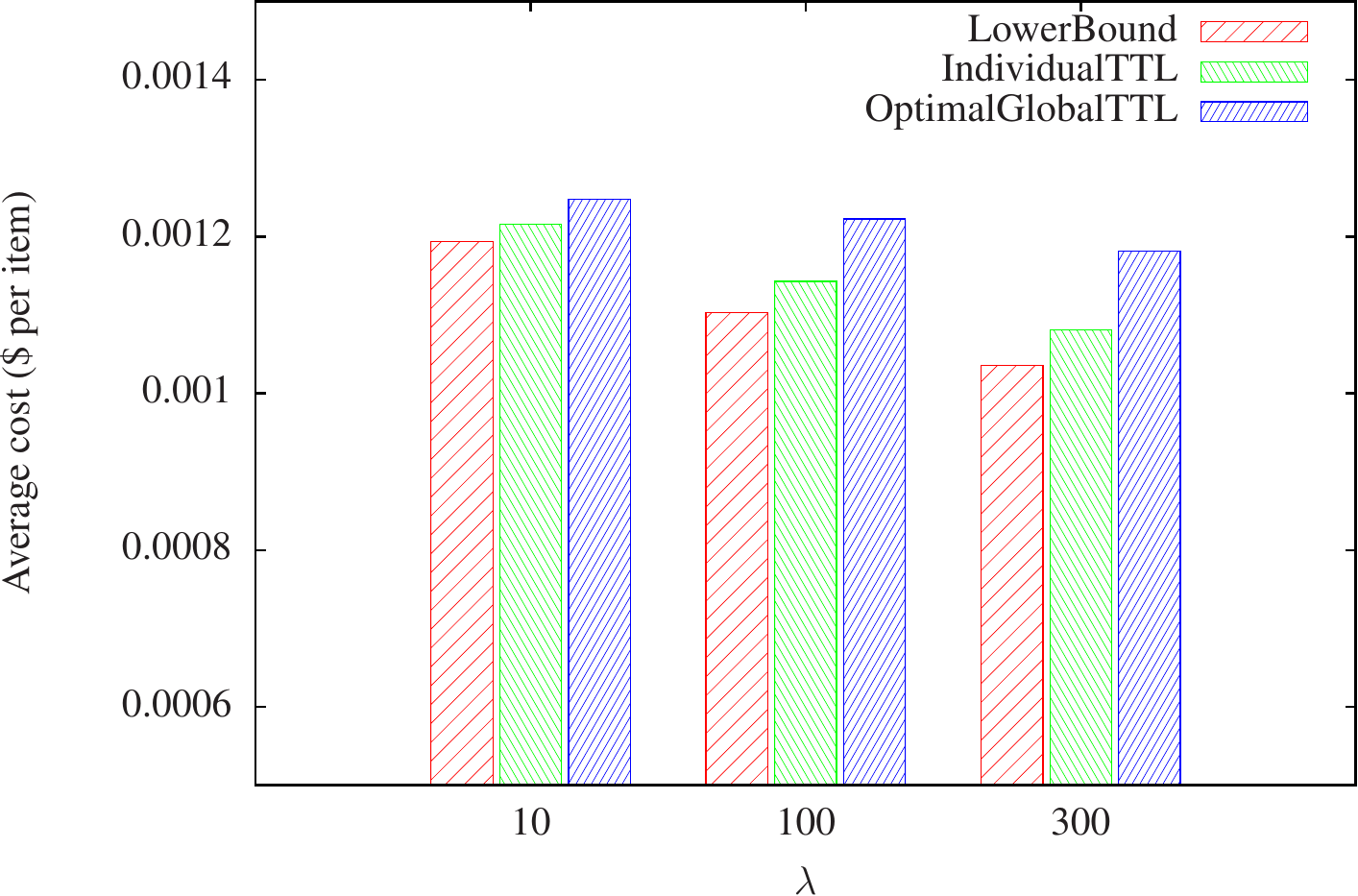}
\caption{Cache policies cost per item with Poisson arrivals.}
\label{fig:PerfComp}
\end{center}
\end{figure}

Figure~\ref{fig:PerfComp} plots the cost per item for each policy. {\em Individual TTL} systematically performs better than {\em Global TTL} and approaches the {\em Lower bound} policy.

\subsubsection{Using trace based simulation}
\begin{figure}
\begin{center}
\includegraphics[width=0.75\columnwidth]{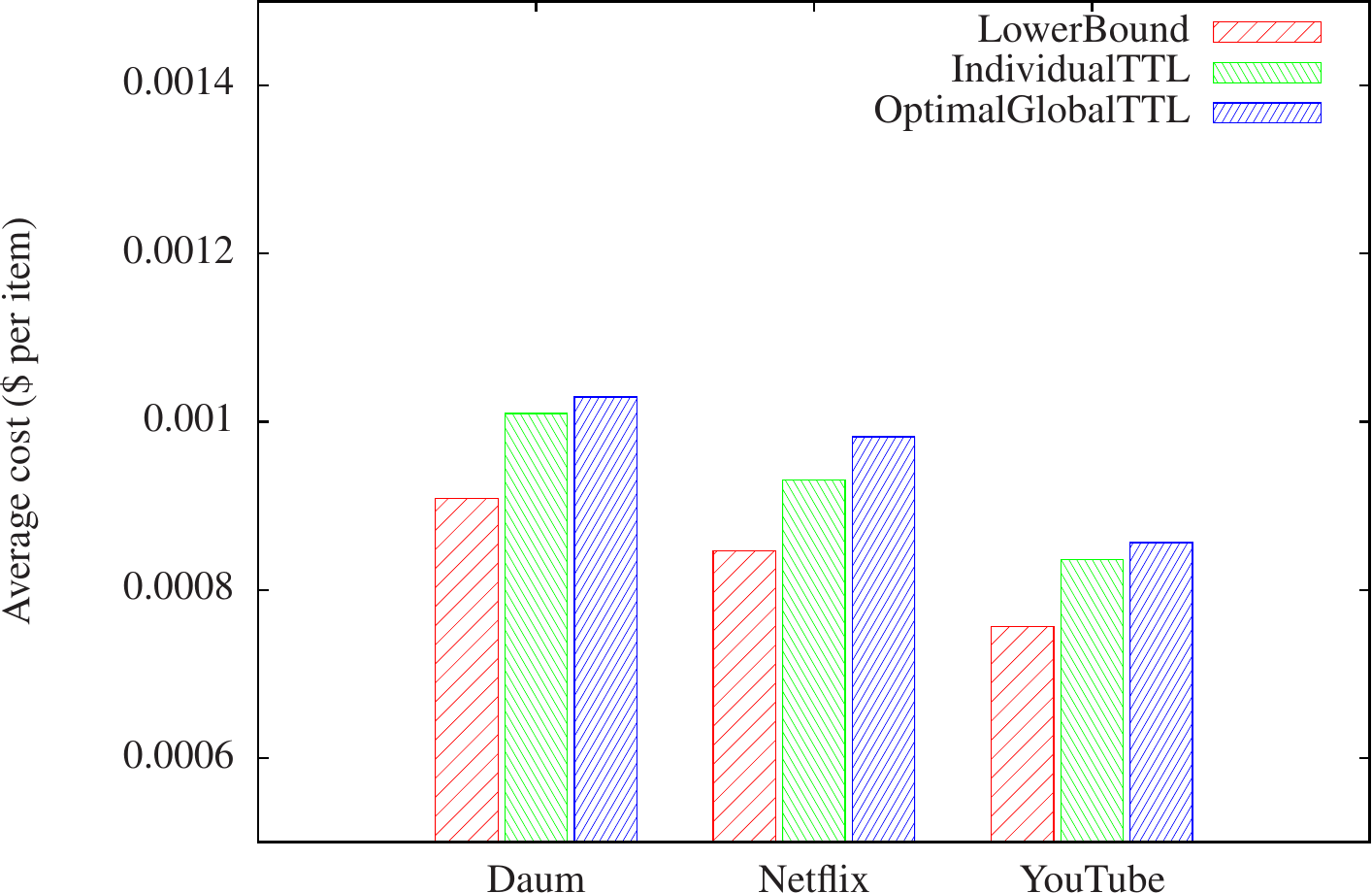}
\caption{Cache policies cost per item on Daum, Netflix and YouTube traces.}
\label{fig:TracePerfComp}
\end{center}
\end{figure}

We now evaluate the cost per item of {\em Global TTL}, {\em Individual TTL} and {\em Lower bound} policies using traces from Netflix, YouTube and Daum.
For the {\em Global TTL} policy we determine the TTL value that minimizes the cost for this policy. We approximate this TTL value by computing the global arrival rate for a trace and searching for the TTL value that minimizes Equation~\eqref{eq:Global TTL}. The optimal TTL values for the {\em Global TTL} policy are $100$, $275$ and $300$ hours for Daum, Netflix and YouTube respectively. 

Figure~\ref{fig:TracePerfComp} plots the cost per item for the considered traces and policies. Although real traces do not follow exponential laws, the {\em Individual TTL} policy outperforms the {\em Global TTL} policy for the three traces.  {\em Individual TTL} policy provides therefore two main advantages : \emph{(i)} it is practical to deploy as {\em Individual TTL} can be inferred using a sliding window and does not require an a priori knowledge of the user request arrival rate and content and ad popularity; \emph{(ii)} it provides better performances than optimal {\em Global TTL} or LRU based policies (which require a priori knowledge of system parameters for optimizing the cache size or the TTL).

\subsection{Sliding Window Size of Individual TTL Policy}
\label{sec:sliding}
We now evaluate the impact of the sliding window duration on the {\em Individual TTL} caching policy. We can notice, that a sliding window with a duration $d$ can measure arrival rates greater than $1/d$ but cannot measure arrival rates smaller than $1/d$. Since we want to determine if an arrival rate is greater or smaller than $\frac{S}{C}$, we need a sliding window with a duration of at least $\frac{C}{S}$.

\begin{figure}
\begin{center}
\subfigure[Using synthetic simulation]{
\includegraphics[width=0.8\columnwidth]{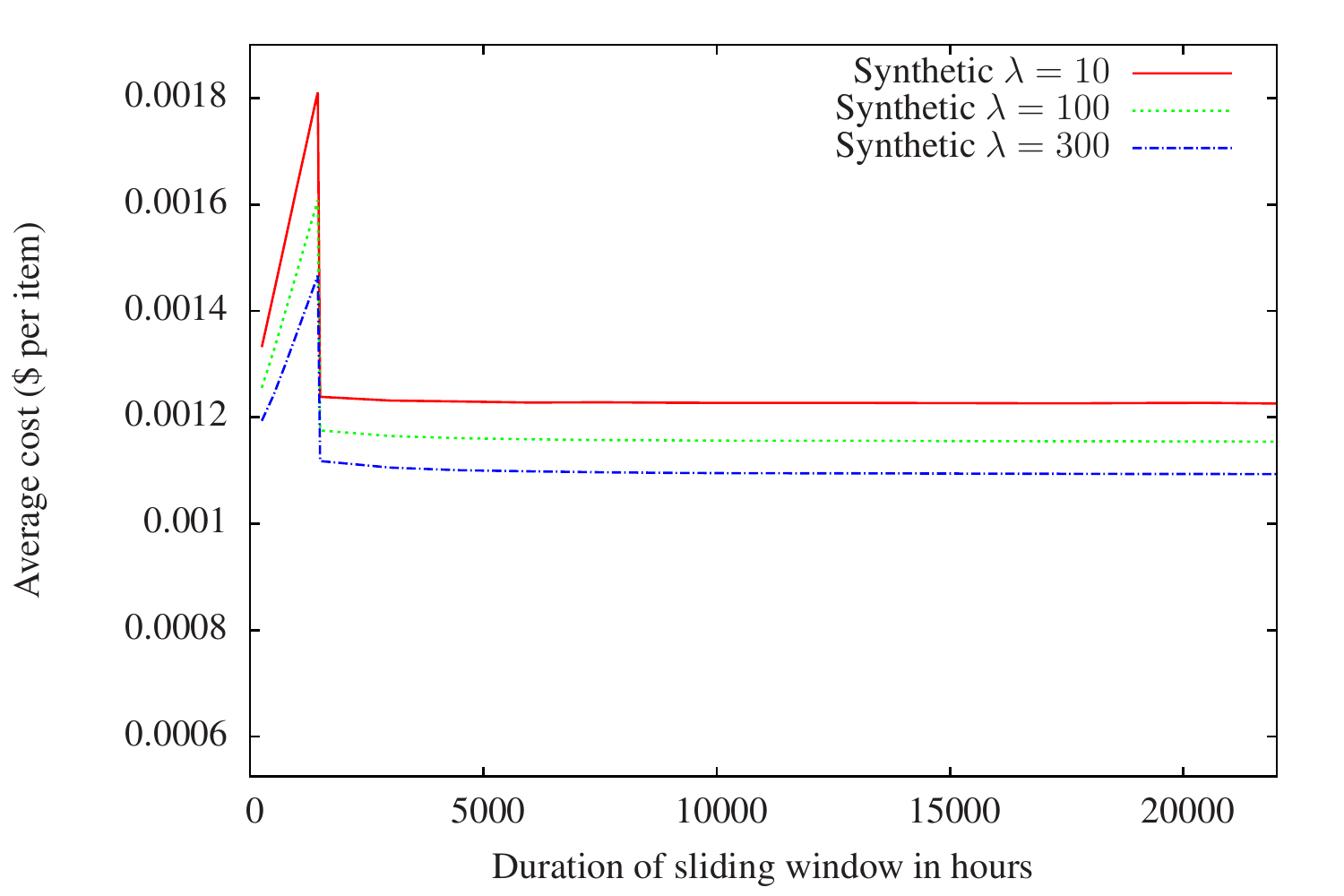}
\label{fig:WindowSizeSynth}
} %
\subfigure[Using trace-based simulation]{
\includegraphics[width=0.8\columnwidth]{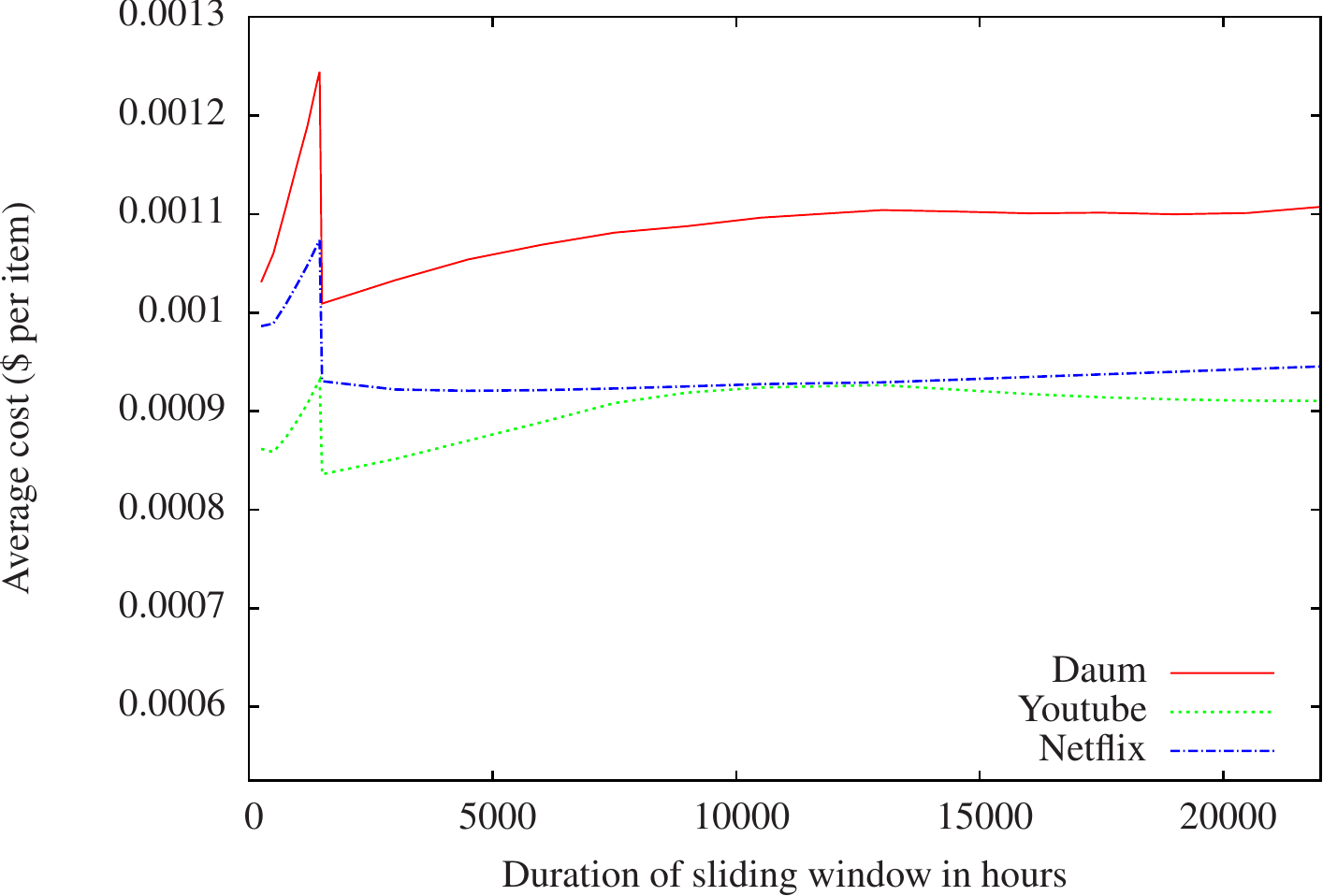}
\label{fig:WindowSizeTrace}
} %
\caption{Mean cost per item as a function of the duration of the sliding window using {\em Individual TTL}.}
\label{fig:WindowSize}
\end{center}
\end{figure}

Figure~\ref{fig:WindowSizeSynth} and Figure~\ref{fig:WindowSizeTrace} show the cost per item as a function of the sliding window duration using the {\em Individual TTL} cache policy using synthetic simulation and trace-based simulation respectively.
We can notice a clear drop in cost around $1481$ hours which corresponds to $\frac{C}{S}$ in all traces and synthetic simulation, which validates that sliding window durations less than $\frac{C}{S}$ are not adequate.

After $\frac{C}{S}$, with synthetic simulation (Figure~\ref{fig:WindowSizeSynth}) the cost per item quickly converges and then stays constant. This is not surprising since the arrival rate for a given item is constant with the simulation. I.e. the estimation becomes more precise with an increasing sliding window duration. Yet, the cost decrease slowly after $\frac{C}{S}$ making this sliding window width a good trade-off between computational efficiency and precision.

\begin{newtext}
After $\frac{C}{S}$, with trace-based simulation (Figure~\ref{fig:WindowSizeTrace}) the different traces behave slightly differently. For Netflix, we observe a slight decrease in cost until $2\frac{C}{S}$ before it increases again; the estimation of instantaneous arrivals becomes less precise when the cost increases. With Daum and YouTube, there is a clear minimum at $\frac{C}{S}$ and then a general trend of cost increase. Indeed, in realistic traces, contrary to synthetic traces, the viewing rate of a movie is not constant: a movie may become popular for some period (\emph{e.g.}, for one month after the release of the movie, for a few days after a newspaper links to a video). Daum and Youtube must have many video that become popular for a short period (less than $\frac{C}{S}$ which is equal to 2 months). On Netflix, movies probably remain active for a bit longer explaining the optimal window of 4 months.
\end{newtext}

\begin{newtext}
The above evaluation highlights that the most adequate window duration is not the widest possible but $\frac{C}{S}$. For some cases (\emph{e.g.}, Daum, Youtube), it corresponds to a minimum. For other cases (\emph{e.g.}, Netflix, synthetic traces), it provides a good approximation while keeping the computational costs low.
\end{newtext}

\section{Related Work}
\label{ref:related}

Most related work rely on the LRU or LFU caching policies or some flavor of these two policies. 
\cite{Cao1997} studies HTTP proxy caching policies, considering parameters such as the item size or the cost to retrieve the item from its origin. \cite{Rizzo2000} proposes an LRU flavor that considers the cost and benefit of purging an item from an HTTP proxy cache. 
ARC \cite{Megiddo2003} is a flavor of LRU with uniform page sizes, applicable to hard-disk drives or CPU caches. SOPA \cite{Wang2010} dynamically chooses a policy among a set of policies 
according to the observed workload.
Time based policies have only been considered for maintaining data consistency \cite{Liu1997, Sit2002}, as detailed in Section~\ref{sec:background}. 

Many of today's cloud systems use {\em memcached} \cite{petrovic2008using, gordon2011low, vaquero2011dynamically}, a distributed caching system that maintains data objects in RAM. It is typically used to speed up access to database items. Memcached is limited by the RAM size and employs an LRU policy to select the data items maintained in RAM.

Chiu et al. \cite{Chiu2010} propose to cache and share in the memory of dedicated EC2 compute instances the results of multiple cloud-based services. 
In contrast to memcached, the cache automatically scales up and down as a function of the demand. 
EC2 instances are allocated and deallocated according to the required cache size. The authors propose an item eviction strategy based on a sliding window with a decay factor for older item accesses. Chiu et al. exclusively rely on cloud compute and do not try to optimize cloud costs.

Yuan et al. \cite{yuan2010cost} study the trade-off of cloud compute versus cloud storage in scientific workflows. Scientific workflows handle large amount of data and generate intermediate data necessary to perform the scientific computation. The intermediate data may either be recomputed each time or stored. The authors build an intermediate data dependency graph and identify the intermediate data that is worth storing in the cloud instead of recomputing it each time. Yuan et al. do not consider request arrival rates. While the general approach of trading cloud compute versus cloud storage is similar to our approach, their proposed model is specific to the considered application. Their approach does not fit a more general caching system. 

Banditwattanawong et al. \cite{Banditwattanawong2012} study a caching policy for a proxy deployed in a consumer premise and aiming at reducing public cloud data-out traffic. The cache policies only consider fixed size architectures and take into account cloud data-out traffic cost to decide which item to keep in cache. 

\section{Conclusion}
\label{ref:conclusion}

We proposed caching policies tailored to the specificities of cloud-based caching, which impose no limit on the cache capacity and adopt a pay-per-use cost model. The proposed caching policies therefore depart from classical caching policies such as LRU that manage a fixed capacity. The objective of the proposed caching policies is to reduce the cloud cost.

\begin{newtext}
We provided analytical models for time-based caching and derived a caching policy, {\em Individual TTL}, that is easy to operate, since it does not require any a priori knowledge of the user arrival rates or movie popularities. {\em Individual TTL} minimizes the cost individually for each item, based on the observed item features (size, request rate, storage and compute costs). Other items do not impact the caching decisions for one particular item. The caching policy adapts over time to evolving request rates or prices.
We validated our analytical model and evaluated our caching policies using synthetic and trace-based simulation. 

The results presented in this paper can apply broadly to systems where there exist a cost for storing and for generating an item on a pay-per-use basis with unbounded resources.
As a perspective, it could be interesting to perform a similar work on other aspects of systems that have been studied with the assumption that resources are fixed or bounded, and to reconsider the design decisions that have been taken. This can lead to simpler or more efficient systems.
\end{newtext}

\bibliographystyle{abbrv}
\bibliography{biblio}

\end{document}